# Modulation of spin dynamics in a channel of a non-ballistic spin field effect transistor.

Ehud Shafir[a], Min Shen[a] and Semion Saikin[a,b*]
[a]Center for Quantum Device Technology

Department of Physics
Department of Electrical and Computer Engineering
Clarkson University, Potsdam, NY 13699, USA

[b]Department of Physics, Kazan State University, Kazan 420008, Russia

[*]email: saikin@clarkson.edu

*Abstract.*

We have investigated the effect of the gate voltage on spin relaxation in an $Al_{0.3}Ga_{0.7}As/GaAs/Al_{0.3}Ga_{0.7}As$ heterostructure. The study is motivated by a recent proposal for a non-ballistic spin field effect transistor that utilizes the interplay between the Rashba and the Dresselhaus spin-orbit interaction in the device channel. The model, which utilizes real material parameters, in order to calculate spin dynamics as a function of the gate voltage has been developed. From the obtained results we define the efficiency of the spin polarization and spin density modulation. The estimated modulation of the spin polarization at room temperature is of the order of 15-20%. The results show that the effect is not sufficient for device applications. However, it can be observed experimentally by optical pulse-probe techniques.

Successful applications utilizing giant magnetoresistance and tunneling magnetoresistance effects in layered ferromagnetic-metal structures for commercial devices [1] have motivated large interest in spin dependent phenomena in semiconductor structures [2,3]. In comparison with metal based structures, semiconductor spintronic devices are belief to be compatible with conventional circuits and more flexible in functionality. Many semiconductor spin electronic (spintronic) devices have been proposed recently [4-14]. The optimistic expectation is that such devices could be scalable to smaller sizes, dissipate less power in comparison to conventional devices and in addition utilize the property of spin quantum coherence [15-17]. According to a more skeptic estimation semiconductor spintronic devices will be limited only to specific applications [18,19]. In order to clarify this controversy the functionality of the various proposed device structures should be analyzed.

We have estimated the efficiency of the gate voltage control over the spin polarization in a channel of a non-ballistic spintronic field effect transistor (spin-FET) [10] at room temperature utilizing realistic material parameters. This spin-FET should be stable to effects of electron scattering, in contrast to other spintronic devices that operate in a ballistic transport regime. It utilizes spin relaxation of conduction electrons in III-V or II-VI semiconductor quantum wells (QW) modulated by the gate voltage [10]. In such structures the spin dynamics of the conduction electrons is controlled by the spin-orbit interaction [20,21]. Two different spin-orbit terms are present in zincblende semiconductor heterostructures, Rashba term [20]

$$H_\text{R} = \eta(k_y\sigma_x - k_x\sigma_y), \tag{1}$$

and Dresselhaus term [22]

$$H_\text{D} = \gamma(k_y\sigma_y - k_x\sigma_x). \tag{2}$$

The interplay between these terms makes the spin relaxation strongly anisotropic [23]. It was shown theoretically that for some particular configurations spin polarized electrons can be transported without substantial loss of polarization if the spin-orbit coupling constants $\eta$ and $\gamma$ are nearly equal [10, 24]. The external gate voltage controls the difference between the Rashba and Dresselhaus terms (mostly through the variation of the Rashba term [26,27]). As a result, it produces a different spin polarization at the drain contact. The spin-dependent filtering of the ferromagnetic drain contact [4] converts the spin polarized current to charge current.

In this study we have addressed the issue of spin polarization and spin density modulation by controlling the gate voltage, without taking into account the issues of injection and filtering. It should be noted that recent experimental advances allow efficient electrical spin injection [28] and spin detection [29] at room temperature.

The structure considered in this work is shown in Fig. 1. We have assumed that the thermalized spin polarized electrons are injected into the QW at the left boundary, $x = 0$.

The device channel ($x$ axis) is oriented along the (1 -1 0) crystallographic direction and initial spin polarization is parallel to (-1 1 0) direction. Within a drift-diffusion approximation [25,30] the electron spin density at a given position $x$ will be [25]

$$n_s(x) = n_s(0)e^{-x/L_s}. \tag{3}$$

The characteristic spin dephasing length, $L_s$, is defined as

$$L_s = \left( \frac{\mu E}{2D} + \sqrt{\left(\frac{\mu E}{2D}\right)^2 + \left(\frac{2m^*(\eta-\gamma)}{\hbar^2}\right)^2} \right)^{-1}, \tag{4}$$

where $E$ is the in-plane electric field, $\mu$ is the electron mobility and $D$ is the diffusion coefficient. Within this model [25], spin polarization is conserved along the channel if the spin-orbit coupling constants are equal and the electric field is negative. On the other hand, it decays exponentially if they are different. The effect of a finite electric field in Eq. (4) is accounted similarly to [31].

A simulation model has been developed based on [32] to compute the properties of the heterostrucutre shown in Fig.1. The algorithm consists of coupled macroscopic and microscopic models. The macroscopic model which includes the transport and continuity equations is solved iteratively with Poisson equation to get the self-consistent electron concentration and electric potential in the whole device. The total number of electrons in the QW is obtained from the concentration distribution. To take into account the quantum effect of the QW, the microscopic model, Schrödinger equation, is solved self-consistently with Poisson equation. It refines the distribution of the concentrations in subbands and the potential in the region of the device where the quantum mechanical effects take place (QW width plus 5nm in both sides out of the QW). The exact form of the energy band diagram of the heterointerface is based on the Anderson model [33,34]. The resulting electric potential distribution, conduction band profile, subband energies and confined electron wavefunctions are used to calculate the Rashba and the Dresselhaus terms as functions of the gate voltage and device structure.

In the simulation, the gate-semiconductor interface was assumed to be a Schottky barrier contact and the substrate interface was assumed to be an ohmic contact. The boundary conditions of the device are represented by the voltage and the concentration, which were calculated using the equations from [35]. The Schottky barrier height, $\phi_B = 1.06$ eV, was obtained from experimental data [36]. In the QW region the three lowest subbands were accounted. Profiles of doping density and material composition are assumed to vary only in one dimension. Effects of the crystal potential were parameterized by an effective mass, which is constant in each material region, and which changes abruptly at the material interface. The Schrödinger equation, solved for the electrons confined in the QW, is spin independent. That is applicable for middle gap semiconductor structures [37].

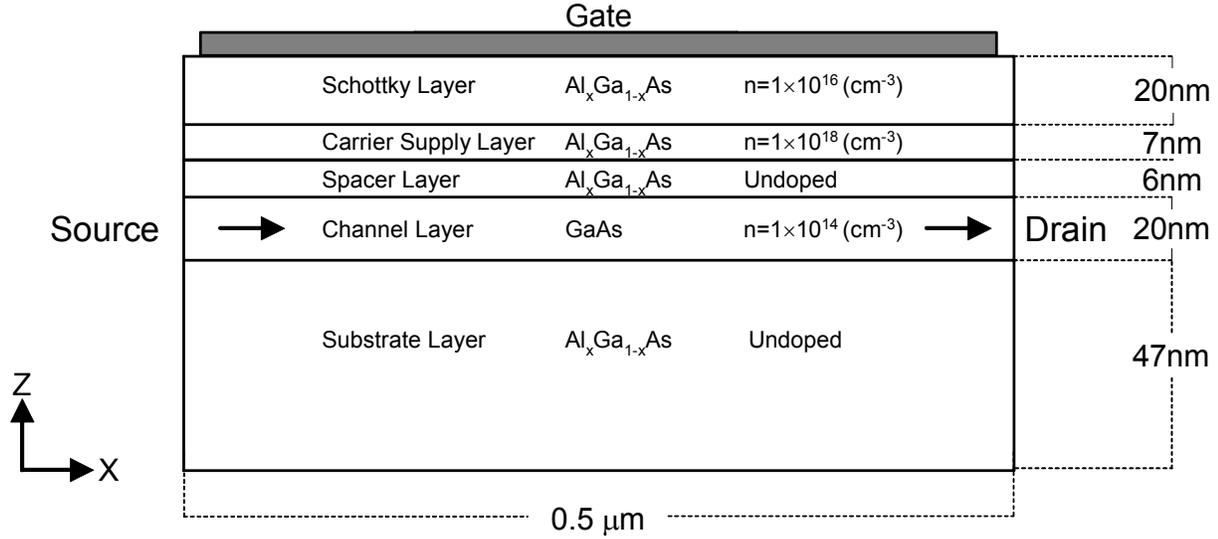

**Fig. 1.** Schematic diagram of the structure simulated, x = 0.3.

We have calculated the spin orbit coefficients for two cases. The first case (Case I) is based on the simplified equations, $\eta = \alpha \overline{E}$ (Rashba constant is independent on the subband index *i*) and $\gamma_i = \beta \langle k_z^{i^2} \rangle$, where, $\alpha = 5.33$ eVÅ² and $\beta = 29$ eVÅ³ were fitted to the experimental data [27]. $\overline{E}$ is the average electric field in the QW region and $\langle k_z^{i^2} \rangle$ is the expectation value of the wave vector squared. In the second case (Case II) we used the model developed by W. Zawadzki and P. Pfeffer [37]. The utilized constants were $E_0(\text{GaAs}) = -1.424$ eV, $E_0(\text{AlGaAs}) = -1.798$ eV, $\Delta_0(\text{GaAs}) = -0.34$ eV, $\Delta_0(\text{AlGaAs}) = -0.328$ eV [38], $E_1(\text{GaAs}) = 3.04$ eV, $E_1(\text{AlGaAs}) = 2.693$ eV [39], $P_0 = 10.493$ eVÅ, $P_1 = 4.780$ eVÅ, $Q = 8.165$ eVÅ, $\overline{\Delta} = -0.050$ eV [40].

The calculated spin-orbit coupling coefficients, presented in the Fig. 2, are compatible with the results of experimental measurements and theoretical calculations [26,27,41-43]. According to Case I, Fig 2a, at $V_g = 0.8$ V the Rashba and Dresselhaus coefficient are equal. In Case II, Fig 2b, this occurs at a higher gate voltage that was out of the studied range.

It can be seen from the results in Fig. 2b that the Rashba coefficient is linear and nearly the same for all the subbands. In both cases $\eta = 0$ at $V_g = 0.42$V, where the QW is symmetric. The slope of $\gamma_i$ in both figures can be explained as the effect of the electric field on $k_z$. It is more pronounced in the lowest subband, which is highly sensitive to the conduction band variations.

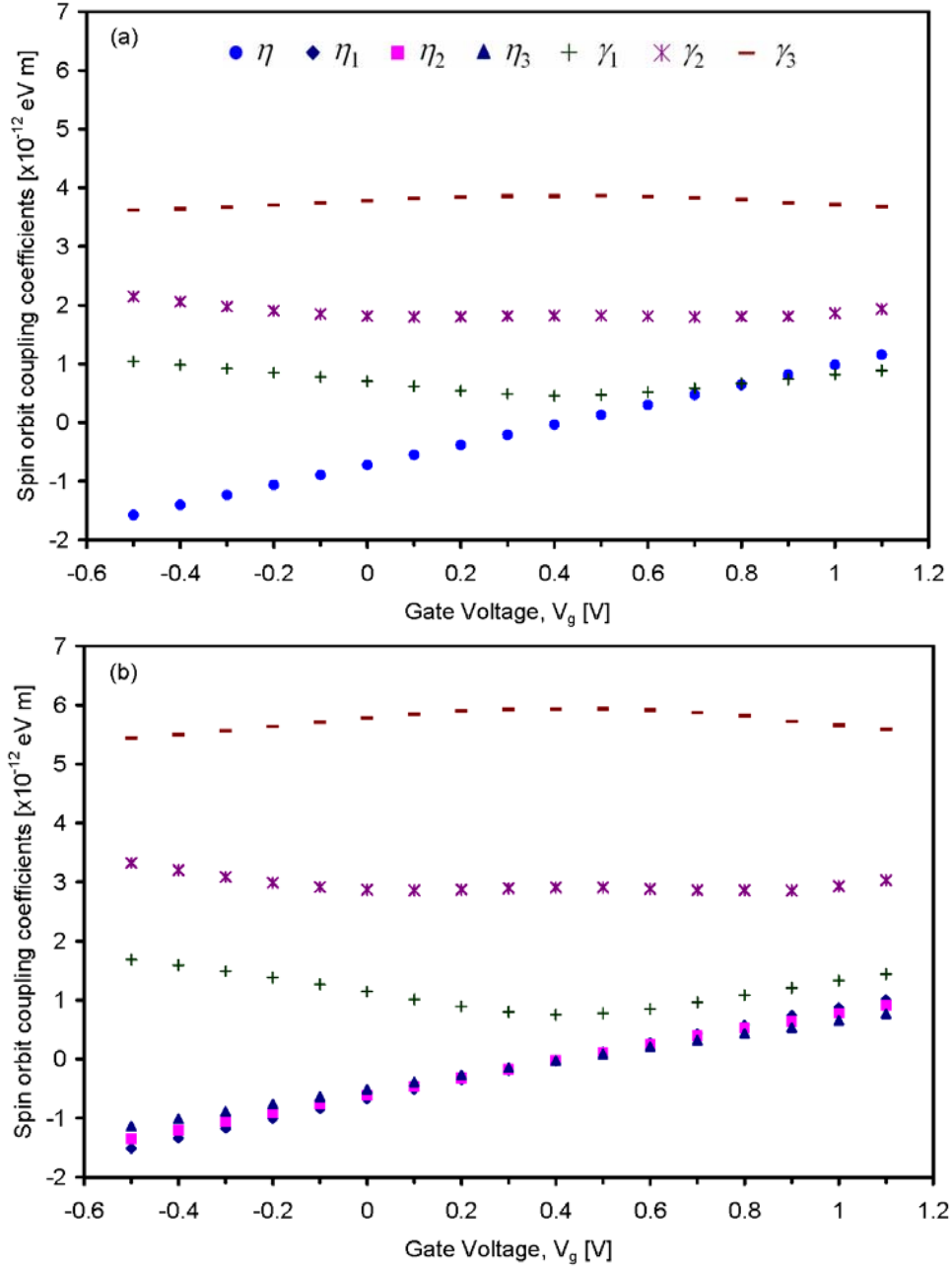

**Fig. 2**. Rashba, $\eta_i$, and Dresselhaus, $\gamma_i$, spin-orbit coupling coefficients as a function of the gate voltage. Indexes correspond to different subbands. (a) Case I – phenomenological parameters fitted to experimental data [27], (b) Case II – parameters calculated according to [37].

It can be deduced from the results that it will not be possible to control electron spin dynamics over all subbands simultaneously. Hence, in order to achieve a bigger spin modulation, the majority of the electrons should be transported on the first subband. This corresponds to the conclusions by [44].

To characterize the efficiency of the spin density modulation we introduce the ratio between spin densities at the drain contact, $x = a$, for two different values of gate voltages

$$\Gamma = \frac{\sum_i n_s^i(a, V_g^{off})}{\sum_i n_s^i(a, V_g^{on})}. \tag{5}$$

$V_g^{off}$ and $V_g^{on}$ correspond to "off" and "on" states of the spin-FET and $i$ is the subband index. The parameter $\Gamma$ in Eq. (5) characterizes two different processes. The first is the conventional charge manipulation as in FET. The second is the modulation of the spin polarization, $P = n_\sigma / n$. Equation

$$\Lambda_P = 1 - \Gamma / \Gamma_n \tag{6}$$

defines the range of the spin polarization variation. In the same form as (5) we define $\Gamma_n$ as the ratio between the average electron charge concentrations, $n^i(a, V_g^{off})$ and $n^i(a, V_g^{on})$.

To estimate the parameters $\Gamma$ and $\Lambda_P$ the in-plane electric filed is specified as $E = -10^5$ V/m. The "on" and "off" states of the gate voltage are defined as $V_g^{on} = 0.4$ V and $V_g^{off} = 0.3$ V respectively. The device length and the applied voltage range correspond to trends in semiconductor devices [45]. The calculated values, Table I, show that the efficiency of the gate modulation is not sufficient for a device application. However, variation of spin polarization can be observed experimentally at room temperature using optical measurement techniques [46].

Similar results for $\Lambda_P$ were obtained for an inverted structure, in which the QW is located between the gate contact and carrier supply layer.

**Table 1**. Efficiency of the spin density modulation, $\Gamma$, and range of spin polarization variation $\Lambda_P$. The results are given for two different sets of spin-orbit coefficients. Case I – phenomenological parameters fitted to experimental data [27]. Case II – parameters calculated according to [37]

Table I

|  | Case I | Case II |
|---|---|---|
| $\Gamma$ [%] | 12 | 11.4 |
| $\Lambda_P$ [%] | 15.2 | 19.8 |

*In conclusion*. We have studied the effects of the gate voltage on spin relaxation in a spin-FET device using realistic parameters. The spin dynamics in the structure is governed by the spin orbit interaction. The calculations of the spin-orbit coupling coefficients for a given device structure and gate voltage are based on the self-consistent steady state solution of the Poisson and Schrödinger equations coupled with the transport and continuity equations.

We have shown that it is possible to control electron spin relaxation in a quantum well by the gate voltage at room temperature. The calculated range of the spin polarization modulation for the "on" and "off" states is approximately $\Lambda_P$ =15-20%. The effect can be measured using optical pulse-probe techniques. In order to make it applicable for commercial devices further improvements of the structure design should are required.

*Acknowledgements*. We are grateful to Ming-Cheng Cheng, Vladimir Privman and Israel Vagner for useful discussions. This research was supported by the National Security Agency and Advanced Research and Development Activity under Army Research Office contract DAAD-19-02-1-0035, and by the National Science Foundation, grant DMR-0121146.